\documentstyle[aaspp4,flushrt]{article}
\input epsf

\def\et{{\it et \thinspace al.}\ }

\def\cm3{cm$^{-3}$}

\def\spose#1{\hbox to 0pt{#1\hss}}

\begin{document}

\title{PHYSICAL PARAMETERS OF ERUPTING LUMINOUS BLUE VARIABLES:
NGC\,2363-V1 CAUGHT IN THE ACT\altaffilmark{1}}

\author{
Laurent Drissen$^2$, 
Paul A. Crowther$^3$, 
Linda J. Smith$^3$,
Carmelle Robert$^2$,
Jean-Ren\'{e} Roy$^2$
and D. John Hillier$^4$}

\affil{(2) D\'{e}partement de Physique, 
Universit\'e Laval and Observatoire du
mont M\'egantic\\Québec QC G1K 7P4, Canada\\
(3)Department of Physics and Astronomy, University College London\\
Gower Street, London, WC1E 6BT, UK\\
(4)Department of Physics and Astronomy, University of Pittsburgh\\
Pittsburgh, PA~15260}

\altaffiltext{1}{Based on observations with the NASA/ESA
Hubble Space Telescope, obtained at the Space Telescope Science
Institute, which is operated by AURA, Inc., under NASA contract NAS5-26555.}

\begin{abstract}
A quantitative study of the Luminous Blue Variable NGC\,2363-V1 
in the Magellanic galaxy NGC\,2366 (D = 3.44 Mpc)
is presented, based on ultraviolet and optical 
{\it Hubble Space Telescope} STIS spectroscopy. Contemporary WFPC2 and 
William Herschel Telescope imaging reveals a modest V-band brightness 
increase of $\sim$0.2 mag per year between 1996 January--1997 November, 
reaching V=17.4 mag, corresponding to $M_{\rm V}$=$-$10.4 mag. Subsequently,
V1 underwent a similar decrease in V-band brightness, together with a UV
brightening of 0.35 mag from 1997 November to 1999 November. 

The optical spectrum of V1 is  dominated by H emission lines, with 
Fe\,{\sc ii}, He\,{\sc i} and Na\,{\sc i} also detected. In  the ultraviolet, 
a forest of Fe absorption features and 
numerous absorption lines typical of mid-B supergiants (such as Si\,{\sc ii},
Si\,{\sc iii}, Si\,{\sc iv}, C\,{\sc iii}, C\,{\sc iv}) are observed. 
From a spectral analysis with the 
non-LTE, line-blanketed code of Hillier \& Miller (1998), we derive 
stellar parameters of $T_{\ast}$=11kK, $R_{\ast}$=420$R_{\odot}$,
log\,$(L/L_{\odot})$=6.35 during 1997 November, and 
$T_{\ast}$=13kK, $R_{\ast}$=315$R_{\odot}$,
log\,$(L/L_{\odot})$=6.4 for 1999 July. The wind properties of V1
are also exceptional, with $\dot{M} \simeq 4.4 \times 10^{-4} M_{\odot}$
yr$^{-1}$
and $v_{\infty}\simeq$300 km\,s$^{-1}$, allowing 
for a clumped wind (filling factor = 0.3), 
and assuming H/He$\sim$4 by number.

The presence of Fe lines in the UV and optical
spectrum of V1 permits an estimate of the heavy elemental abundance 
of NGC\,2363 from our spectral synthesis. Although some deficiencies remain,
allowance for charge exchange reactions in our calculations supports 
a SMC-like metallicity, that has previously been determined for NGC\,2363
from nebular oxygen diagnostics. 

Considering a variety of possible progenitor stars, V1  has definitely 
undergone a giant eruption, with a substantial increase in stellar 
luminosity, radius, and almost certainly mass-loss rate, such that its 
stellar radius increased at an average rate of 
$\sim$4 km\,s$^{-1}$ during 1992 October -- 1995 February. The stellar properties
of V1 are compared to other LBVs, including $\eta$ Car and HD\,5980 
during its brief eruption in 1994 September, the latter 
newly analyzed here. The mass-loss rate of
the HD\,5980 eruptor compares closely with V1, but its bolometric luminosity 
was a factor $\sim$6 times larger.
\end{abstract}


\section{INTRODUCTION}

Luminous Blue variables (LBVs) are evolved massive stars which
display photometric and spectroscopic
variability with different timescales and amplitudes. Although LBVs show
micro-variations ($\Delta$mag $\simeq$ 0.1-0.3 mag; $\Delta$t $\sim$ days-months)
like other supergiants, it is their moderate ``excursions'' 
($\Delta$mag $\simeq$ 1-2 mag at constant $M_{bol}$; $\Delta$t $\sim$ years)
and giant eruptions ($\Delta$mag $>$ 3 mag with an increase in $M_{bol}$; 
$\Delta$t $\sim$ years - decades) that define them as a class. 
They are found in the upper left section of the H-R diagram, having
$M_{bol} \leq -9$ and log ($T_{eff}$) $\geq$ 4.1 (except during an eruption
when their surface temperature often drops to log ($T_{eff}$) $\simeq$ 3.85).
LBVs are often, but not always, surrounded by a small (1-3 pc) nebula 
resulting from a giant eruption. 
We refer the reader to Humphreys \& Davidson (1994) and
Nota \& Lamers (1997) for comprehensive, recent reviews on the subject.
Although it is now widely accepted that the LBV phenomenon represents a normal
yet unstable evolutionary phase in the life of the most massive stars,
the exact mechanism leading to the major 
eruptions is still unknown. This lack of understanding is due in part to the
problems involved in the theoretical modeling of the eruption phenomenon
(Stothers 1999b), but also to the
rarity of such events which has until recently prevented us from determining
the physical parameters of an LBV {\it during} an eruption.

In 1996, we reported the discovery of a bright variable star
in the  giant extragalactic H~\,{\sc ii} region NGC\,2363 (Drissen,
Roy \& Robert 1996, 1997, hereafter DRR97). This star, 
known as NGC\,2363-V1 (hereafter V1), was first noticed in 
HST/WFPC2 images. An archival search for ground-based images allowed us 
to determine that this star was below the detection limit in 1991 and 1992 
($V \geq 21.5$), but became visible in late 1993. In early 1995, it became the 
brightest star in its galaxy, with $V = 18.0$ mag. 
The photometric behavior, absolute 
magnitude  and the presence in its spectrum of a strong H$\alpha$ emission line 
suggested that V1 is an LBV currently experiencing
a major eruption event.

The host of the current LBV event is the giant H\,{\sc ii} region NGC\,2363,
located at the end of the bar in the Magellanic galaxy NGC\,2366 (D = 3.44
Mpc; Tolstoy \et 1995). The metallicity
of NGC 2363 has generally been determined to be close to that of the SMC
(Gonzalez-Delgado \et 1994), but Luridiana, Peimbert \& Leitherer (1999)
suggest that the metallicity of this region has been underestimated, and that a
value of $Z = 0.25 Z_\odot$ would give a better fit to the observed
line ratios.
A detailed study of the stellar content of NGC\,2363 (Drissen \et 2000)
reveals that two clusters of different ages are responsible for the 
ionization of the nebula. The main contributor is a very young (1 Myr 
or less), dense cluster of massive stars still embedded in their natal molecular 
cloud. V1 is a member of the second cluster, which is between 3 and 
5 Myr old and contains 3 Wolf-Rayet stars. 

Given the importance of this star in our understanding of the LBV
phenomenon, we proposed a spectroscopic and photometric follow-up
of V1 with the {\it Hubble Space Telescope}. This project was granted 
the long-term status for cycles 7 to 9 (GO- 7391, 8403 and 8781). 
We present in this paper 
an analysis of HST/STIS spectrograms 
obtained in 1997 November and 1999 July, 
covering the wavelength range 1180 \AA\ to 1$\mu$m, supplemented by 
ground-based and 
HST photometry. We present our new observations in $\S$~2 and 
produce a quantitative
analysis with the non-LTE line blanketed code of Hillier \& Miller (1998) in 
$\S$~3. A plethora of Fe\,{\sc ii} lines in the UV and optical spectrum
of V1 permits an estimate of the heavy metal content for NGC\,2363.
Our results are discussed in $\S$~4, with conclusions reached in $\S$~5.

\section{OBSERVATIONS}\label{obs}

DRR97 published a light curve of NGC\,2363-V1 between 1991--1996
revealing a dramatic increase of $\ge$3.5 mag in brightness, although
no color information was available. 
In this section, we present recent HST/WFPC2 and 
ground-based imaging, plus HST/STIS spectroscopy.

\subsection{Photometry}

UBVRI images of V1 were obtained on the night of 1998 January
14/15 with the 4.2 m William Herschel Telescope (WHT) on La Palma,
Canary Islands. The Cassegrain auxiliary port was used with a 
Tektronix $1024 \times 1024$ CCD ($0.''11$/pixel). The exposure times
were 300 s at BVRI and 600 s at $U$. The seeing was measured to be
$0.9$--$1''.0$, sufficient to resolve V1 from the nearby H\,{\sc ii} region.
Landolt (1992) standard star fields (98 624, 98 626 and 98 634)
were also observed to provide the zero point
calibration for the V1 images. 

Accurate magnitudes for V1 are impossible to obtain using conventional
aperture photometry because of severe contamination from the nearby
H\,{\sc ii} region. The following approach was therefore adopted. Magnitudes
were measured using a 35 pixel aperture with DAOPHOT (Stetson 1987)
for several isolated stars in the same frames as
V1. The same stars and V1 were then measured using a much smaller
aperture of 3.5 pixels, and sky inner and outer radii of 4 and 7
pixels. These values were chosen to minimize the H\,{\sc ii} region
contamination. The UBVRI magnitudes of V1 were then determined by
assuming that the same fraction of light has been
excluded from the smaller apertures for all the measured stars.


We also obtained WFPC2 photometry with the F170W, F336W, F547M and F1042M
filters on 1998 March and December, 1999 November and 2000 February. 
The images were
processed with the standard pipeline and aperture photometry of V1 
was performed with DAOPHOT. The prescriptions given by Whitmore (1995) 
and Holtzman \et (1995) were then followed
to transform the instrumental magnitudes into the Johnson system.

Finally, near-infrared imagery of NGC\,2363 were obtained in 1997 January by
D. Devost and R. Doyon with the Monica Camera attached to the 
Canada-France-Hawaii telescope, which is described in more details in 
Drissen \et (2000). From these observations, we get J=17.6 ($\pm 0.1$),
H = 17.2 ($\pm 0.1$) and K = 16.9 ($\pm 0.2$). These values are
slightly fainter than the predicted magnitudes based on the models
described in the next sections (J = 17.1, H = 16.9 and K = 16.7), indicating that
cold dust has not yet formed in the stellar outflow.

The results of all photometric measurements are given in Table 1.
The agreement between the ground-based and HST March data (taken 6 weeks apart)
is very good in the V band, but the difference in U (0.6 mag) is significant.
An updated light curve with the new data (in the V band)
is shown in Figure~1.  After the original outburst in 1993--1995, 
the brightness increase has been much more
modest, though constant during 1996--1997, with a rate of $\sim 0.2$ mag per
year,
as if the star had reached a state of quasi-equilibrium.
The 1998-2000 photometric observations suggest a small, but comparable,
faintening in the V band.
The WFPC2 data also indicate that V1 brightened by $\sim 0.3$ mag at 170 nm
between 1998 March and 2000 February; during the same period, it faded
by about the same amount at 1 $\mu$m.

\subsection{Spectroscopy}

We obtained HST/STIS spectra of NGC\,2363-V1 in 1997 November and 1999 July, 
using the
MAMA detectors short ward of 3000\AA\ and the CCDs at longer wavelengths.
Because of the sensitivity of the MAMA detectors to high radiation levels
in the South Atlantic Anomaly,
the 1997 optical and UV spectrograms of V1 had to be acquired on separate occasions
three weeks apart (November 2 and 21). Given the relative photometric
stability of V1 between 1997 November and 1998 December, we do not expect
that the spectroscopic appearance of the star has significantly changed during 
this short interval. In 1999, the optical and UV spectra were obtained
2 days apart. The following gratings were used: 
G140L (160 minutes) and G230L (90 minutes) with the MAMA detector;
G430L (50 minute exposure), G750L (35 minutes) and G750M (H$\alpha$, medium
resolution; 50 minutes) with the CCD detector. 
The 0.2 arcsecond-wide slit was used in all cases, 
which corresponds to a linear scale of 3 pc at the distance of NGC\,2366.

Because the data originally processed with the standard HST pipeline 
showed unexplained anomalies, we have retrieved our data from the 
Canadian Astronomical Data Center (CADC). 
The CADC archives uses the best available calibration
files to re-calibrate the data. Moreover, because V1 is located in the
NGC 2363 nebula, extra care has been taken in subtracting the adjacent
nebular spectrum from the spectrum of V1. Nebular contamination is small
(V1 is in the middle of an expanding cavity) and relatively easy to
subtract in the high (spatial) resolution STIS data compared with the highly
contaminated ground-based spectra, but the standard pipeline calibration
did not do a good job in this respect.
Standard IRAF procedures were used. Despite the use of contemporary
flatfield calibrations, the far red portion of the spectra is
contaminated by fringing.
By convolving our observed spectrophotometry with suitable synthetic filters,
we obtained wide-band Johnson photometry (see Table~1) that is in 
excellent agreement with WFPC2 imaging.


The low-resolution flux calibrated HST datasets are presented in 
Figure~2.
Balmer emission lines are the most prominent spectral features in the
visible; weaker lines at He\,{\sc i} $\lambda$4471, 
5876, 6678, Na\,{\sc i} $\lambda$5890
plus numerous Fe\,{\sc ii} transitions are also observed. 
The apparently noisy near-UV depression results from many 
Fe-transitions, rather than high interstellar reddening. 

Since individual features are generally unresolved with
the low resolution gratings, we have also obtained 
higher resolution H$\alpha$ STIS observation with the G750M grating 
for each epoch. These are presented in 
Figure 3 and reveal a very strong H$\alpha$ P Cygni profile betraying the
extent of V1's envelope and its high mass-loss rate.
The observed H$\alpha$ flux is 5.6$\times$10$^{-14}$ and
6.3$\times$10$^{-14}$ erg\,s$^{-1}$\,cm$^{-2}$
for 1997 November and 1999 July, respectively. Correcting for interstellar
reddening (see below) and the distance to NGC\,2363, the 
intrinsic H$\alpha$ luminosity of V1 is 2.4--2.7$\times 10^{4} L_{\odot}$.

While the general appearance of the spectrum is similar at both epochs, 
a few changes are worth noting:

(1) The UV continuum flux (1200 - 2800 \AA ) is stronger in 1999.
However, the line fluxes in this wavelength range
has not changed significantly.

(2) The strength and profile of the Balmer lines has changed (see
Table 2). The flux of the H$\alpha$ line in the July 99 dataset 
is $\sim$20\% stronger and its 
blue absorption component extends to higher velocities (Figure 3).
On the other hand, the equivalent widths of the other Balmer lines have 
slightly decreased and the absorption components of their P Cygni profile 
are less deep in 1999.

(3) The FeII lines at 4925 and 5016 \AA\ which were strong
in the 1997 data became much weaker in 1999.

\subsection{Morphology of the UV spectrum}

The resolution of the STIS data ($\sim$ 300 km s$^{-1}$) does not allow us 
to easily, and unequivocally, distinguish between the interstellar 
and purely stellar contribution components to the resonance lines on the 
basis of their width alone. But a direct comparison of the line strengths 
between the spectrum of V1 and those of knots NGC 2363-A and B (obtained 
with HST/FOS; see Drissen \et 2000 and Figure 4 here) should allow us to 
identify the lines which are intrinsic to the star's photosphere.
We assume that since V1 and knot B are both located within the same 
expanding superbubble, the spectral contribution from their local ISM
will be similar, although knot B is a cluster of OB stars and
some ISM lines in its spectrum could be contaminated by a stellar 
contribution. 
On the other hand, the ISM local to knot A could be different, but 
contributions from OB stars are negligible. The contribution from Milky 
Way gas should be the same in all three objects. 
From Figure 4, which shows the rectified spectrum of V1 in 1999
in the range 1160 - 1700 \AA\ along with that of knots A and B, and  from
Table 3, which lists the equivalent width of the most conspicuous 
UV lines in the three objects, we conclude that the UV lines of V1 have
a strong photospheric contribution. As examples, one may compare the
lines of Si\,{\sc ii} $\lambda$1260, C\,{\sc ii} $\lambda$1335, Si\,{\sc iv} $\lambda$1400 and Si\,{\sc ii} $\lambda$1527.
None of these UV lines in V1 however show any evidence for an emission
component, suggesting a rather low effective temperature.

The following lines are both strong in V1 and absent or very weak in
the spectrum of knots A and B; they are thus likely to have
a photospheric origin: C\,{\sc iii} $\lambda$1175, Mg\,{\sc ii} $\lambda$1240, 
S\,{\sc ii} $\lambda$1251,54,60,
Si\,{\sc ii} $\lambda$1265, Ti\,{\sc iii}+Si\,{\sc iii} between 1286 and
1303\AA , 
Si\,{\sc ii} $\lambda$1309, and Si\,{\sc ii} $\lambda$1533.

The C\,{\sc iii} $\lambda$1175 line typically shows a strong P Cygni profile
in late-type O to early-type B supergiants and becomes mostly photospheric
in mid B supergiants (Snow \& Morton 1976). It is an obvious
($W_\lambda = 2.4$ \AA ) absorption feature in V1, but appears much 
weaker than in late-type B supergiants such as
$\eta$ CMa (B5Ia) or $\beta$ Ori A (B8Ia).
Also, features seen in the spectrum of V1 between 1295 
and 1303 \AA , are often associated to Si\,{\sc iii} in B stars (Massa 1989).
Therefore in Figure 4, we have plotted along with V1 the rectified spectrum
of an average Galactic B5.5I star (based on high resolution IUE spectra 
of 4 B6I and B5I stars from the library of Robert 1999).
The resolution of this B5.5I spectrum was degraded to mimic the STIS data.
Similarities with V1 are striking.  In the following discussion we
compare the spectral feature of V1 with those of B stars. 

The strength of the Si\,{\sc ii} $\lambda$1265 line is a strong (inverse) 
function of the effective temperature in OB stars, but insensitive to the
luminosity class (Massa 1989; Prinja 1990), 
increasing from 100 m\AA\ at $T_{eff} = 25 000$K to 1600 m\AA\ 
at $T_{eff} = 13 000$K. Si\,{\sc ii} $\lambda$1265 is not seen in stars hotter than B0.
The large
value of $W_\lambda$ (Si\,{\sc ii} $\lambda$1265) = 1800 m\AA\ in the
1999 spectrum of V1 sets a relatively stringent upper limit of 
$T_{eff} \leq 13 000$K, albeit based on a Galactic supergiant calibration.

The strong line at $\sim$ 1303 \AA\ is a blend of stellar and 
interstellar lines. The interstellar component (mostly O\,{\sc i} $\lambda$
1302 and Si\,{\sc ii} $\lambda$1304) dominates in knots A and B (total 
$W_\lambda$ = 2.2 \AA\ in both cases), but the blend is stronger in V1
($W_\lambda$ = 4 \AA ), indicating a strong photospheric contribution
from Si\,{\sc iii} $\lambda$ 1301,03 and Si\,{\sc ii} $\lambda$1304
typical of late to mid-B stars.

The Si\,{\sc ii} $\lambda$1309 line behaves like Si\,{\sc ii} $\lambda$1265 as
a function of OB spectral types. 
The Si\,{\sc ii} $\lambda$1533 is more difficult to study in hot OB stars
as the broad wind profiles developing in C\,{\sc iv} $\lambda$1550 
engulfs it and as
it is blended with Fe\,{\sc iv} (Nemry \et 1991) in hotter stars.
Nevertheless there are indications that Si\,{\sc ii} $\lambda$1533 
is also a good B star signature.

Spectral features around 1317, 1345, and 1455 \AA , which are
associated with Ni\,{\sc ii}, P\,{\sc iii}, and Ti\,{\sc iii}, 
are typical of B supergiants; they are not seen in 
giants and dwarfs B (except maybe in the earliest types B0-1). 

The strength of the doublet Si\,{\sc iv} $\lambda$1394,1402 is strongly temperature
and luminosity dependent (Walborn \et 1995). It is absent from all late B
luminosity classes,
appears in absorption in mid-early types, and in the case of the early B 
supergiants shows a strong P Cygni wind profile.
An absorption doublet of C\,{\sc iv} $\lambda$1549,51 
is only seen in hot B0-2 dwarfs and giants.
A wind profile appears in the hottest giant B0.
In the case of supergiants, the absoption is seen for mid B and
quickly becomes a strong P Cygni profile in type B4 and hotter.
We do not see Si\,{\sc iv} and C\,{\sc iv} wind profiles in V1
but rather absorption typical of B5-6 I stars.

Around 1600 \AA\, we see many strong lines of Fe\,{\sc ii} and Fe\,{\sc iii}
which resemble
the features of B supergiants. These lines are much weaker in 
B giants and dwarfs.

We therefore conclude that the UV spectrum of V1 is similar to that of
a B supergiant star of type 5-6.

\section{SPECTROSCOPIC ANALYSIS}\label{analysis}

An important consideration for the interpretation of V1's
spectral energy distribution is the determination
of the extinction. Gonzalez-Delgado \et (1994)
derived E(B-V)$\simeq$0.15 mag from a ground-based
integrated nebular spectrum of the whole H~\,{\sc ii} region. Since the
extinction is likely to be different in knot A, which dominates
the global nebular spectrum, than in the middle of the expanding
bubble where V1 and knot B lie, we have extracted the spectrum of the
nebula from the pixels adjacent to V1 in the STIS long-slit spectroscopy.
The nebular H$\alpha$/H$\beta$ line strengths from this region reveal
a low value of E(B-V)=0.06 using Case~B recombination theory 
(Hummer \& Storey 1995); this is
in perfect agreement with Drissen et al. (2000) who recently estimated 
E(B-V)=0.04 (Galaxy) + 0.02 (NGC\,2363) based on HST/FOC UV spectra
of knot B in NGC 2363. 
Therefore, if an associated circumstellar nebula is present, it does not
add measureably to the interstellar extinction, arguing against a
dense, $\eta$ Car-like nebula surrounding V1.
We have thus reddened the theoretical
continuum energy distributions by E(B-V)=0.04 using the Seaton (1979) 
Galactic extinction law, plus  E(B-V)=0.02 using the Bouchet et al. (1985) SMC
law. Assuming a distance of 3.44Mpc to NGC\,2366 implies
$M_{\rm V}$=$-$10.44 mag during 1997 November, and $-$10.29 in 1999 July.

For stars with extended atmospheres such as NGC\,2363-V1, as evidenced from
its P Cygni H$\alpha$ profile, the usual 
assumptions of plane parallel geometry and local thermodynamic equilibrium 
(LTE) are totally inadequate.  In this section we introduce the atmospheric
model that we shall use for V1 and discuss our analysis technique.

\subsection{Atmosphere code}

We utilize the iterative, line blanketed  technique 
of Hillier \& Miller (1998)  which solves the 
transfer equation in the co-moving frame subject 
to statistical and radiative equilibrium in an 
expanding, steady-state atmosphere.
Populations and ionization
structure are consistent with the radiation field. Line blanketing is
treated via a global Doppler line width of $V_{\rm Dop}$=10 km\,s$^{-1}$.

Despite the high quality dataset available, the spectroscopic 
analysis of NGC\,2363--V1 has proved to be difficult, principally because 
its low excitation
state means that Fe\,{\sc ii} is the dominant blanketing ion, which is extremely
complex. In addition, as discussed by Hillier et al. (1998), mass-loss and
hydrogen content are highly coupled. Since we are unable to tightly constrain 
the stellar temperature of V1, due to the absence of lines from adjacent
ionization stages of the same element, we cannot derive a mass-loss rate 
without first fixing its He abundance. Consequently, we shall adopt an abundance
of H/He=4 by number, which is typical of LBVs (Crowther 1997), and subsequently 
investigate the effect of different He abundances on the mass-loss rate.
Similarly, metal abundances were fixed at 0.20$Z_{\odot}$ as indicated by
Gonzalez-Delgado et al. (1994) for NGC\,2363, and varied once stellar parameters were
derived.

A simplifying `super level' approach 
is used for individual levels (Anderson 1989), particularly for iron-group 
elements. In this approach, several levels of similar energies and properties
are treated as a single `super level', with only the populations of
the super level included in the solution of the rate equations. The
population of an individual atomic level in the full model atom is
determined by assuming that it has the same departure coefficient as 
the corresponding super level to which it belongs.

The present model include representative model atoms for H, He, N, 
Mg, Al, Na, Ca, Si and Fe, as indicated in Table~4, in which a total of 
32,191 transitions are considered. Details of each ion are included, 
such that for hydrogen, $n$=1 to 30 full levels are considered, 
which are grouped into 10 super levels. Our default atomic dataset
for Fe\,{\sc ii} was taken from Nahar (1995), in which 827 atomic levels,
were combined into 134 super levels, producing 21,907 transitions (22,924
weak transitions with log $gf<-4$ were omitted). 
Hillier et al. (2000) have recently analyzed the UV and optical spectrum 
of $\eta$ Car based on the Nahar compilation and found improved comparisons
with observation if Kurucz (1993) oscillator strengths were selected instead
of those from Nahar, although negligible differences were obtained for V1.

In the outer wind of V1, and other LBVs, hydrogen becomes neutral,
so it is necessary to allow for charge exchange reactions. Of particular relevance to LBVs is the 
charge exchange reaction:
Fe$^{2+}$ 3d$^{6} (^{5}$D) + H $\leftrightarrow$ Fe$^{+}$ 
3d$^{6}$4s ($a^{6}$D) + H$^{+}$ 
(Neufeld \& Dalgarno 1987)
since the ratio of charge recombination to radiative recombination
for Fe$^{2+}$ is $\sim 10^{3}$ H/H$^{+}$ (Hillier et al. 2000). 
This has a major influence on the ionization structure, and 
significantly enhances 
the Fe\,{\sc ii} line strengths. Also relevant to LBVs is the 
nitrogen reaction:
N$^{2+}$ + He$^{+}$ $\leftrightarrow$ N$^{+}$ + He$^{2+}$ 
(Herrero et al. 1995).

Finally,  since it is well established that
the dense winds of early-type stars are clumped 
(e.g. L\'epine \& Moffat 1999; Eversberg et al. 1998),
we consider a relatively simple volume filling factor 
approach for V1 following Hillier \& Miller (1998). 
Estimates of clumping factors in the optical line forming 
region for V1 are provided by simultaneously matching recombination 
emission line strengths (sensitive to the square of the density) 
and electron
scattering wings (sensitive to density) as discussed by Hillier (1991).

\subsection{Stellar parameters}

Since we were unable to use lines from adjacent ionization stages of
one particular element to derive stellar temperatures, we instead relied
on the stellar continuum, de-reddened as described above.
We have calculated a grid of models for V1 at each epoch in order to 
simultaneously match the stellar continuum distribution from which
the stellar temperature and luminosity are derived, plus H$\alpha$
which provides an excellent tracer of mass-loss in the winds of 
early-type stars (Puls et al. 1996). We have adopted a typical H/He content 
of LBVs for NGC\,2363-V1, namely H/He=4 by number (see also below). 

For the 1997 November dataset, we find optimum agreement for a model
in which $T_{\ast}$=11kK, log($L/L_{\odot}$)=6.35 and $R(\tau_{\rm Ross}=10)$
= $420 R_{\odot}$, as shown in Figure 5 (upper panels). 
Since the atmosphere of NGC\,2363-V1 is very extended, 
$R(\tau_{\rm Ross}=2/3)$ = 590$R_{\odot}$, the formal `effective 
temperature' is $T_{\rm eff}$=9,200K. Similar properties for July 1999
are $T_{\ast}$=13kK, log($L/L_{\odot}$)=6.4 and $R(\tau_{\rm Ross}=10)$
= $315 R_{\odot}$, shown in Figure 5 (lower panels). Therefore, the 
spectroscopic differences between the two epochs represent variations
in stellar radius (i.e. temperature) at fairly constant bolometric luminosity.
This naturally explains 
the simultaneous UV brightening and optical fading during this period. 

While the agreement between the model and the observations is generally
very good, we wish to highlight several important discrepancies revealed in
this work. The sole helium line
observed at high resolution with HST/STIS is 
He\,{\sc i} $\lambda$6678, which is
observed in absorption for both epochs. It was hoped that we would have been
able to determine H/He abundances from H$\alpha$ and He\,{\sc i}, but 
the He\,{\sc i} absorption is rather insensitive to the precise 
H/He abundance ratio, and more problematic, predicted absorption strengths
are significantly weaker than observations.

Another important discrepancy is that the continuum fit to observations
is imperfect, at both epochs; (a) Fe\,{\sc ii} blanketing is 
apparently underestimated between $\lambda\lambda$1700-2300 
using our determination of the iron content in V1 (see $\S$~3.4); (b) 
the models clearly fail to
reproduce the observed continuum distribution either side of the 
Balmer `jump', by up to $\sim$20\%. Hillier et al. (2000) experienced
similar difficulties in their analysis of $\eta$ Car (see $\S$~3.5).

\subsection{Mass-loss properties}

In addition to deriving stellar parameters, we have simultaneously 
modeled the H$\alpha$ profile for each epoch by varying the 
mass-loss rate, and terminal velocity. 

Using volume filling factors of $\sim$30\%, which provide a reasonable match 
to the electron scattering wings of H$\alpha$, we find an approximately
constant $\dot{M}$=4.4$\times 10^{-4} M_{\odot}$ yr$^{-1}$  for both epochs, 
such that the stronger H$\alpha$ emission in 1999 results from the increase
in stellar temperature, rather than a change 
in mass-loss rate. Wind velocities are also very similar, 
$v_{\infty}$=325 km\,s$^{-1}$ and 
$v_{\infty}$=290 km\,s$^{-1}$ for 1997 November and 1999 July, respectively.
Differences in H$\alpha$ absorption can be matched with different turbulent
velocities, $v_{\rm turb}$=10 and 40 km\,s$^{-1}$.
Comparisons between synthetic spectra and observations are shown in Figure
~6. 

An important caveat which should be kept in mind is that we have relied solely
on H$\alpha$ as a mass-loss indicator, since 
all other Balmer lines are unresolved. 
Nevertheless, low dispersion HST/STIS observations reveal that 
the increase in strength of H$\alpha$ from 1997 to 1999 
is not repeated for H$\beta$ and higher Balmer members (Table~2). Were we 
instead to rely on the H$\beta$ flux, our modeling would
reveal a $\sim$30\% decrease in mass-loss, rather than a constant rate, between
the two epochs. Such differences in behavior amongst members of the 
Balmer series are not repeated amongst other
LBV-type supergiants (e.g. HDE\,316285: Hillier et al. 1998).

One further complication is that since we 
have used solely H$\alpha$ to constrain the mass-loss rate, this is
also dependent on the atmospheric H/He composition. For a solar composition
the clumped mass-loss rates would decrease 
to $\sim2.5\times 10^{-4} M_{\odot}$
yr$^{-1}$. Conversely, if V1 is more highly enriched (e.g. H/He=1 by number), 
we would need to increase the mass-loss rate estimate by a factor of $\sim$2
to $10^{-3} M_{\odot}$ yr$^{-1}$.

\subsection{Metal Abundances}

Unfortunately the low spectral resolution of our observations together
with the narrow, weak emission from most metal ions means that the
determination of elemental abundances is not possible. However, 
the presence of numerous optical Fe\,{\sc ii} transitions, together with
a forest of metal lines in the near-UV permit an estimate of the
iron abundance to be determined. Neglecting the effect of charge exchange
would require systematically higher abundances (at least by a factor of two)
than we obtain here. Ideally, we would like to have used a similar star
at known metallicity to check our modeling procedure against, but no such
stars exist with parameters close enough to V1. 

In Figure 7 we compare de-reddened ultraviolet observations of V1 from 1997 November
with Fe\,{\sc ii} blanketed models at 0.20$Z_{\odot}$, 
0.5$Z_{\odot}$ and 1.0$Z_{\odot}$ degraded to the resolution of the 
STIS spectroscopy. Each model provides a fair match to the observed
continuum depression around 2500\AA\, including the discontinuity 
at $\lambda$2630. The 0.20$Z_{\odot}$ case compares very closely with
observations in the $\lambda\lambda$2500--2900 region. The 0.10$Z_{\odot}$
model, not shown in this Figure, is marginally worse than
the 0.20$Z_{\odot}$ case. Discrepancies 
remain for each case, especially at $\lambda\lambda$1750--2300. 
We attribute such discrepancies to the shear complexity of the observed 
UV spectrum (Hillier et al. 2000 obtained similar problems for 
models of $\eta$ Car). For example, in the case of the 1.0$Z_{\odot}$ 
model, 2878 lines are predicted to have equivalent widths of 
$\ge$0.1\AA\ in the 1200--3100\AA\ region. These are dominated by 
2351 lines of Fe\,{\sc ii} and 416 lines of Fe\,{\sc iii}. In the light
of model discrepancies between $\lambda$1750--2300, a comparison 
with observations marginally favors the low metallicity cases, but is 
unable to provide a definitive abundance determination.

Figure 8 compares rectified observations in the 3000--5500\AA\ region, with 
synthetic spectra covering the same range in metallicity. Aside from the
Balmer series, the majority of features are again due to Fe\,{\sc ii},
the strengths of which are very sensitive to metal content. As for
the UV, an SMC-like iron abundance for V1 is favored by the weakness 
of iron lines in the $\lambda\lambda$3100--3500, $\lambda\lambda$4450--4600 
and $\lambda\lambda$4900--5300 regions. Unfortunately, prominent P~Cygni 
Na\,{\sc i} emission at $\lambda$5890-6 is not predicted sufficiently strong 
in any case, while comparisons in the far-red are prevented by 
significant CCD fringing.

Note, however, that caution is necessary. Although we include over 700 energy 
levels of Fe\,{\sc ii} in our calculations, this ion is extremely complex,
with potentially imprecise atomic data. In addition, although Fe\,{\sc ii} 
is certainly the most prominent heavy element, ions 
of other less abundant elements will make significant contributions to 
the line blanketing at particular wavelengths (e.g. Ni\,{\sc ii}). 
Work is currently underway to include such elements in our calculations.

Further, we have based our iron abundance estimate on a comparison between 
models selected to match the 1997 November
HST/STIS dataset. Were we instead to have used the 1999 July observation, we
would have revealed a somewhat lower iron abundance, since the emission 
strength of optical Fe\,{\sc ii} lines (e.g. $\lambda$5015) is weaker at
that epoch.

Nevertheless, we conclude that the heavy metal content of 
NGC\,2363-V1 is $\sim$0.2$Z_{\odot}$, in reasonable agreement with the previously 
derived nebular oxygen abundance of 12+log(O/H)=7.9, a factor of six below the 
solar value. Recall that iron is produced by supernovae, whose progenitors 
have lifetimes of typically 1Gyr, while
oxygen is created exclusively by short-lived, high mass stars. Our study 
represents the  most distant star whose iron abundance has been derived, 
and compared with $\alpha$-elements. 

\subsection{Balmer jump -- evidence for asymmetry?}

The shape of the Balmer jump is notoriously difficult to reproduce in 
extreme OB supergiants. The case for V1 is no exception -- 
from Figure 5 the model does not fit
the continuum well on both sides of the Balmer jump. In fact, the
model predicts a lower continuum level shortward of the Balmer edge
while the observations clearly show an increasing flux below 3700 \AA .
This is best illustrated for V1 in Figure 9, and includes 
HST/FOS spectra of two B stars in the
LMC cluster Breysacher 73 (Walborn \et 1999). The spectral shape of
V1 around the Balmer edge is quite similar to that of Brey 73-1C, 
an O9.5-B1pe star, classified as such because an emission component is 
clearly seen in the center of the Balmer absorption lines. The
unusual flux increase from 3750 to 3650 \AA\, which is 
seen in both stars,
has been observed in a number of Be stars (Kaiser 1987) and has been
interpreted as evidence for the existence of a disk-like
extended envelope of low density around the stars. The mass-loss rate
of V1 is much higher than that of Be stars and the two-component model
may not be applicable in this case, but the peculiar shape of its 
continuum around the Balmer edge may indicate that its envelope is
not spherically symmetric. Indeed, most nebulae around LBVs exhibit
axisymmetric morphologies that could have been shaped by non-spherical
mass loss (Nota \et 1995). Spectropolarimetry of the H$\alpha$ line
and its adjacent continuum could be used to measure the degree of
asymmetry, if any, of V1's envelope.

\section{DISCUSSION}\label{discussion}

Our results place tight constraints on the current stellar properties of
NGC\,2363-V1. Major LBV eruptions are thought to coincide with an increase 
in the bolometric luminosity of the star and probably also its mass-loss 
rate. We now consider whether V1 has undergone a genuine eruption, and
subsequently compare its properties with other LBVs, notably HD\,5980
during its brief eruption in 1994 (Barba \& Niemela 1994).

\subsection{Is NGC\,2363-V1 undergoing a genuine eruption?}

Since the presence of NGC\,2363-V1 was unknown prior to its
recent brightening, no spectroscopic or photometric information
is available to us in order to readily constrain its pre-outburst
properties. 
Consequently, we consider three possibilities for its
spectral type prior to 1993, namely an early O, late O, or early B supergiant.

For O3 and O8 supergiant cases, we adopt a 
temperature and bolometric correction scale from Vacca et al. (1996), together
with mass-loss rate estimates following the wind momentum--luminosity 
relation of Puls et al. (1996) for low metallicity (mixture of LMC
and SMC) stars. For the early B supergiant case, we consider the 
possibility that V1 was a dormant, hot LBV, equivalent to AG Car 
or  R127 during their WN11 phase (e.g. Smith et al. 1994), 
whose wind properties are estimated from Crowther (1997).
Properties derived in this way are presented in Table~5.

We first consider the possibility that V1 was a hot, dormant LBV
immediately prior to 1993.
In this case, the low bolometric correction would indicate that the
luminosity of V1 would have been  amongst the
lowest of all known LBVs with $\log (L/L_{\odot}) \leq 5.5$, comparable
with   R110 (Stahl et al. 1990). In contrast, those LBVs which have been know 
to experience a very hot WN11 phase have luminosities in excess of 
$\log (L/L_{\odot}) \geq 6$. Nevertheless, for such a scenario, the
increase in bolometric luminosity corresponds to a factor of 
approximately ten, with a mass-loss rate 
probably increasing by several hundred fold. We neglect the 
possibility that V1 was a cooler, P Cygni-type dormant LBV since 
its bolometric luminosity would then have been unrealistically low.

Alternatively, V1 may have advanced immediately from an O supergiant
to an erupting LBV. The age of NGC\,2363-B, from which V1 is thought
to have originated is 3--5Myr (Drissen et al. 2000), so an 
O8 supergiant progenitor 
represents a realistic possibility. In such a case, the increase in 
luminosity by V1 is less severe (see Table~5), albeit a factor of 
four (or greater), while  the increase in mass-loss rate is perhaps
one thousand fold if the progenitor has wind characteristics comparable
to Magellanic Cloud supergiants. The absence of cool dust around V1
favors an O supergiant progenitor rather than a dormant LBV.

The final possibility, is that V1
advanced directly from an O3 supergiant. These very young ($\leq$2--2.5\,Myr) 
stars have the highest bolometric corrections of any hot massive star, prior to
the Wolf-Rayet phase. We consider this possibility highly unlikely, given
their youth relative to NGC\,2363-B. Nevertheless, for completeness,
the change in stellar and wind properties would be the least severe, 
with potentially only a modest increase in luminosity, together with 
a factor of one hundred increase in mass-loss rate. 

%
%

The rapid brightening of V1 also indicates a 
corresponding increase in the physical size of the star. Based upon 
a late O supergiant progenitor, the surface of V1 expanded from a radius of 
$\sim20$ to 420$R_{\odot}$ in a  little over two years (1992 October
to 1995 February), corresponding to an average rate of +4 km\,s$^{-1}$! 
This calculation assumes that the stellar temperature of V1 in 
early 1995 was comparable to that of 1997 November, which appears 
realistic considering the similarity in photometry (Table~1). Subsequently,
between 1997 November and 1999 July, the radius of V1 decreased
from 420 to 315$R_{\odot}$, at an average rate of $-$1.4 km\,s$^{-1}$. 
These changes support some form of pulsational or dynamical instability 
as the cause of the present eruption (e.g. Guzik et al. 1999; Stothers 1999a). 
Calculations in our group are underway to assess such possibilities.

In summary,  it appears that V1 is a moderately massive star which
underwent an dramatic increase in luminosity, radius and mass-loss rate 
between late 1992 and early 1995, with a probable late O-type 
supergiant or perhaps
hot, dormant LBV progenitor. Estimating an initial mass for V1 is 
extremely difficult since this would rely on a comparison between the 
luminosity of the star prior to eruption with low metallicity evolutionary
tracks (Meynet \et 1994). Nevertheless, an age of 4--5 Myr suggests an initial
40--60$M_{\odot}$ star, with a present luminosity of $\log(L/L_{\odot})=
5.8-6.0$, consistent with a late-type O supergiant precursor.

Consequently, we suggest that the initial and current mass 
of NGC\,2363-V1 is no more extreme than many other LBVs, such as AG Car. 
In contrast, the initial mass of the Pistol star appears to be 
remarkably high ($\ge$200$M_{\odot}$, according to Figer et al. 1998). 

\subsection{Comparison between V1 and other LBVs}

We have shown that V1 is currently undergoing a giant eruption,
and that its stellar and wind properties are exceptional. 
What distinguishes V1 from `dormant' LBVs, such as AG Car?
How do these compare with HD\,5980 which briefly underwent an 
eruption in a similar low metallicity environment during 1994?

The well known Galactic LBV AG Car is currently undergoing huge 
variations in temperature (from 8,000--25,000K) over timescales of 
decades. Its stellar properties have been studied by 
Leitherer et al. (1994) who found a fairly (homogeneous) mass-loss 
rate of $\sim$10$^{-5}$$M_{\odot}$yr$^{-1}$ at a constant luminosity of
log ($L/L_{\odot})$=6.1. Consequently, V1 is currently little more
than a factor of two times more luminous than AG Car, which is surprisingly
small. In contrast, the mass-loss rate of V1 is perhaps 50 times higher,
and appears to be the principal defining characteristic (see below).

Are the properties of V1 during its present eruption typical of other
massive stars? This question is particularly relevant to stellar evolution
models of massive stars in which, brief episodes of very intensive 
mass-loss are predicted or assumed. 
To address this question, we have investigated the stellar properties of 
HD\,5980 during its brief eruption (Barba \& Niemela 1994) based on the
optical spectroscopy obtained by Heydari-Malayeri et al. (1997) during
the visual maximum in 1994 September when V=8.6 mag (Albert Jones,
private communication).
Based on an interstellar reddening of E(B-V)=0.07 mag towards HD\,5980
(Crowther 2000), together with an SMC distance modulus of 18.9 mag
(Westerlund 1997), the absolute visual magnitude of the erupting star
was $M_{\rm V}=-$10.5 mag (all other components play a negligible 
contribution to the continuum flux at this epoch). This is remarkably
similar to the current absolute visual magnitude of V1. However, the
spectral appearance of the HD\,5980 eruptor was considerably different
from V1 at visual maximum. Heydari-Malayeri et al. (1997) revealed 
a much hotter, WN11-type appearance (Fig.~10) so it was intrinsically much
more luminous than V1. 

We have analysed the optical spectrum of HD\,5980 
from Heydari-Malayeri et al. (1997), using the 
Hillier \& Miller (1998) code. Based 
on the usual He\,{\sc i} $\lambda$4471, He\,{\sc ii} $\lambda$4686 and 
H$\beta$ diagnostics, our derived parameters are listed in Table~4, revealing 
that the luminosity of the HD\,5980 eruptor exceeded both V1 and $\eta$ Car 
by a factor 3--6, respectively. Indeed the only other LBV whose luminosity may
compare favourably with HD\,5980 is the Pistol star (Figer  et al. 1998),
although analysis of it is severely hampered by the high 
reddening plus lack of definitive temperature constraints.
A comparison between our new results for HD\,5980 with those of Koenigsberger
et al. (1998) based on observations from 1994 December (when it had 
visually faded by 1.6 mag) suggests that its bolometric luminosity had 
decreased by a factor of five over three months! Note also the discrepant 
H/He abundance pattern resulting from analysis of the two datasets in Table~6. 

The mass-loss rate for HD\,5980 during its brief eruption compares very 
closely with V1. It appears that `giant eruptions' of LBV's in low
metallicity environments exhibit  representative mass-loss rates approaching 
$\sim$10$^{-3} M_{\odot}$yr$^{-1}$. These differ greatly from the 
sole source of mass-loss rate produced in giant LBV eruptions known to 
date,  namely the famous Homunculus produced
by $\eta$ Car between 1837--1960. During this eruption, 
$\sim2-3M_{\odot}$ were ejected (Davidson  \& Humphreys 1997) corresponding 
to $\sim$10$^{-1} M_{\odot}$yr$^{-1}$. Thus far, $\eta$ Car appears to 
be unique in exhibiting such high rates of mass-loss, namely 
 during its 1840's eruption. Indeed, although not presently 
in outburst, $\eta$ Car remains a very extreme
star, as illustrated in Figure 10, with a present luminosity and mass-loss
rate somewhat higher than V1 (see also Table~6). Hence, the
present V1 outburst should not be considered an equivalent to the great
 $\eta$ Car 1837--1860 eruption.

\section{CONCLUSIONS}\label{conclusions}

From our quantitative analysis of an HST/STIS spectrogram of NGC\,2363-V1,
and a photometric follow-up of this star since the beginning of its
present eruption, we conclude as follows.
V1 is a Luminous Blue Variable currently experiencing a major
eruption. Its immediate progenitor star was probably an initial $\approx$60$ 
M_{\odot}$ late O supergiant.

Its physical parameters are exceptional: log ($L/L_\odot$) = 6.35,
T$_\ast$ = 11kK, $R_{\ast}=420 R_{\odot}$ for 1997 November; and
log ($L/L_\odot$) = 6.4, T$_\ast$ = 13kK, $R_{\ast}=315 R_{\odot}$ for 
1999 July. Therefore, the bolometric luminosity of V1 is exceeded by
only, in ascending order, $\eta$ Car (log 
$L/L_\odot$ = 6.7, Hillier et al. 2000), the Pistol star (log 
$L/L_\odot$ = 6.9$\pm$0.3, Figer et al. 1998) and HD\,5980 during 
eruption (log $L/L_\odot$ = 7.1).

Analysis of high
resolution H$\alpha$ observations indicate that the wind is moderately 
clumped, with $\dot{M}\simeq 4.5\times$10$^{-4}M_{\odot}$ yr$^{-1}$ 
and $v_{\infty}\simeq$300 km\,s$^{-1}$. During its brief eruption in 
1994, HD\,5980 possessed a similar mass-loss rate.  Therefore, the mass-loss
rate of $\sim10^{-1} M_{\odot}$ yr$^{-1}$  exhibited by $\eta$ Car during 
its giant eruption in 1840's (Humphreys \& Davidson 1994) is thus far unique. 

The presence of many iron features in the spectrum permits an estimate
of its heavy metal content. Although discrepancies remain, especially at
UV wavelengths, our study reveals a SMC-like Fe content for NGC\,2363-V1, 
which  is in good agreement with previously derived oxygen abundances for 
NGC\,2363. Consequently, the heavy 
metal enrichment of NGC\,2363, produced by Type~I Supernovae, whose 
progenitors are believed to have a lifetime of typically 1\,Gyr, appears 
to mimic that of light elements, such as oxygen, 
which are created exclusively by short-lived (10\,Myr) massive stars.

Observationally, spectroscopic and photometric monitoring of V1 and 
other LBVs will continue, while theoretical studies are underway to
investigate the cause of the outbursts experiences by V1 and HD\,5980,
whether pulsationally, dynamically or rotationally driven.

\acknowledgments

We would like to thank Mohammad Heydari-Malayeri for providing the outburst
spectrum of HD\,5980, Albert Jones for providing his photomety of HD\,5980,
and Kris Davidson and his HST/STIS observing team for providing
the spectrum of Eta Carina in Figure 10.
This investigation was funded in part by the Natural Sciences
and Engineering Research Council of Canada, by the Fonds FCAR of
the Government of Qu\'ebec and by Universit\'e Laval (LD, JRR and CR).
PAC acknowledges financial support from the Royal Society.
We are especially grateful to the staff of the now defunct Royal Greenwich 
Observatory for  obtaining service imaging of NGC\,2363. The William Herschel 
Telescope is operated on the  Island of La Palma by the  Isaac Newton Group 
in the Spanish  Observatorio del Roque de los Muchachos of the 
Instituto de Astrofisica de Canarias.

\newpage

\begin{deluxetable}{llcccccc}
\tablenum{1}
\tablecaption{Photometric variability of NGC\,2363-V1 since 1996}
\tablehead{
\colhead{Epoch} &
\colhead{Observatory} &
\colhead{F170W} &
\colhead{U} &
\colhead{B} &
\colhead{V} &
\colhead{R} &
\colhead{F1042M} \\
}
\startdata

1996 Jan   & HST/WFPC2 & ----- & ----- & 17.82 & 17.88 &  ----- &  -----   \\
1997 Nov   & HST/STIS  &17.25 &16.79&17.55 & 17.43 &17.16& ----- \\
1998 Jan   & WHT/aux   & ----- &16.13&17.53 & 17.66 &17.33& ----- \\
1998 Mar   & HST/WFPC2 &17.23 &16.72& -----  & 17.52 & ----- & 17.15 \\
1998 Dec   & HST/WFPC2 &17.09 &16.79& -----  & 17.55 & ----- & 17.20    \\
1999 Jul   & HST/STIS  & 16.83 &16.85&17.72 & 17.58 &17.20 & ----- \\
1999 Nov   & HST/WFPC2 &16.89 &16.89& ----- & 17.80 & ----- & 17.56    \\
2000 Feb   & HST/WFPC2 &16.98 &16.72& ----- & 17.76 & ----- & 17.50    \\

\enddata
\end{deluxetable}

\begin{deluxetable}{crr}
\tablewidth{0pt}
\tablenum{2}
\tablecaption{Equivalent widths of the Balmer lines in V1}
 
\tablehead{
\colhead{Line} &
\colhead{$W_\lambda$ (1997)} &
\colhead{$W_\lambda$ (1999)} \\
}
\startdata
  
H$\alpha$& 250 & 300 \nl
H$\beta$ &  45 & 35 \nl
H$\gamma$ & 15 & 11 \nl
H$\delta$ & 6 & 5 \nl

\enddata
\tablecomments{The equivalent width, in units of \AA , is for the
emission component only.}
\end{deluxetable}

\begin{deluxetable}{lccc}
\tablewidth{0pt}
\tablenum{3}
\tablecaption{Equivalent widths of the the strongest
UV lines in V1 and NGC 2363}
 
\tablehead{
\colhead{Line} &
\colhead{$W_\lambda$ (V1)} &
\colhead{$W_\lambda$ (knot A)} &
\colhead{$W_\lambda$ (knot B)} \\
 & (\AA) & (\AA) & (\AA) \\
}
\startdata
  
C\,{\sc iii} $\lambda$1175 & 2.4 & ----- & ----- \nl
Si\,{\sc ii} $\lambda$1260 & 2.8 & 0.9 & 1.6 \nl
Si\,{\sc ii} $\lambda$1265 & 1.8 & -----  & -----  \nl
O\,{\sc i}/Si\,{\sc ii} $\lambda$1303 & 4.1 & 2.2 & 2.3 \nl
Si\,{\sc ii} $\lambda$1309 & 1.0 & ----- & ----- \nl
C\,{\sc ii} $\lambda$1334 & 2.5 & 1.6 & 1.6 \nl
Si\,{\sc iv} $\lambda$1393 & 2.0 & 0.6 & 0.6 \nl
Si\,{\sc iv} $\lambda$1402 & 2.3 & 0.5 & 1.1 \nl
Si\,{\sc ii} $\lambda$1527 & 2.1 & 0.8 & 0.8 \nl
Si\,{\sc ii} $\lambda$1533 & 1.9 & $\leq$0.2 & $\leq$0.1 \nl
C\,{\sc iv} $\lambda$1549,51 & 2.8 & ----- & 2.1 \nl

\enddata
\end{deluxetable}

\begin{deluxetable}{lrrrl}
\tablewidth{0pt}
\tablenum{4}
\tablecaption{Summary of model atoms used in radiative transfer calculations}
\tablehead{
\colhead{Ion} &
\colhead{$N_{\rm S}$} &
\colhead{$N_{\rm F}$} &
\colhead{$N_{\rm Trans}$} &
\colhead{Details} \\
}
\startdata
H\,{\sc i}    & 10 & 30 & 435 & $n \le$30. \\
He\,{\sc i}   & 27 & 39 & 315 & $n \le$14\\
He\,{\sc ii}  &  5 &  5 &  10 & $n\le$5 \\ 
N\,{\sc i}    & 44 &104 & 855 & $nl \le$5f2F$^{\circ}$    \\
N\,{\sc ii}   & 23 & 41 & 144 & $nl \le$2p3d$^{1}$P$^{\circ}$ \\
N\,{\sc iii}  &  8 &  8 &  11 & $nl \le$2p$^{3}$$^{2}$P$^{\circ}$ \\
Na\,{\sc i}   & 18 & 53 & 507 & $n \le$12 \\
Mg\,{\sc ii}  & 18 & 45 & 362 & $n \le$10 \\
Al\,{\sc ii}  & 26 & 44 & 171 & $nl \le$5d$^{1}$D \\
Al\,{\sc iii} & 12 & 20 &  67 & $n \le$5 \\
Ca\,{\sc ii}  & 17 & 58 & 463 & $n \le$11 \\
Si\,{\sc ii}  & 31 & 49 & 281 & $nl \le$3s$^{2}$9h2H$^{\circ}$ \\
Si\,{\sc iii} & 12 & 20 &  42 & $nl \le$3s4p$^{1}$P$^{\circ}$ \\
Fe\,{\sc ii}  &134 &827 &21907& $nl \le$3d$^5$($^6$S)4p$^{2}$($^3$P)$^{4}$P  \\
Fe\,{\sc iii} & 69 &607 & 6620 & $nl \le$3d$^5$($^{4}$D)6s$^{3}$D\\

\enddata
\tablecomments{Including full levels ($N_{\rm F}$),
super levels ($N_{\rm S}$) and total number of transitions ($N_{\rm Trans}$),
following Hillier \& Miller (1998).}
\end{deluxetable}


\begin{deluxetable}{lllccc}
\tablewidth{0pt}
\tablenum{5}
\tablecaption{Estimates of stellar properties for NGC\,2363-V1 prior to outburst}

\tablehead{
\colhead{Sp. Type} &
\colhead{$T_{\rm eff}$}  &
\colhead{$R_{\ast}$} &
\colhead{log L$_{\ast}$} &
\colhead{log$\dot{M}$} & 
\colhead{$v_{\infty}$} \\ 
 & kK &  $R_{\odot}$ & $L_{\odot}$  & $M_{\odot}$yr$^{-1}$ & km\,s$^{-1}$ \\
}
\startdata

O3\,I      &  50.7   & 18  &  $\le$6.27    &  $-$5.3   &3200 \\
O8\,I      &  35.7   & 22  &  $\le$5.85    &  $-$6.2   &2000 \\
B0.5\,I    &  25.6   & 27  &  $\le$5.46    &  $-$5.8   & 250 \\

\enddata
\tablecomments{We use here the temperature and bolometric correction
scale from Vacca et al. (1996), plus the wind momentum--luminosity
relation of Puls et al. (1996) for low metallicity (LMC and SMC) OB stars,
except for the early B supergiant for which typical LBV wind properties are
shown (Crowther 1997).}
\end{deluxetable}

\newpage

\begin{deluxetable}{lrcccccccl}
\tablewidth{0pt}
\tablenum{6}
\tablecaption{Stellar parameters of V1 and of other well known LBVs}
\tablehead{
\colhead{ Star} &
\colhead{Epoch} &
\colhead{$T$/kK} & 
\colhead{$R/R_{\odot}$} &
\colhead{log L$_{\ast}$}&
\colhead{log$\dot{M}/\sqrt{f}$} &
\colhead{$v_{\infty}$} &
\colhead{H/He} &
\colhead{$M_{\rm V}$} &
\colhead{Reference} \\ 
 & & $\tau_{\rm Ross}$=10 & $\tau_{\rm Ross}$=2/3&$L_{\odot}$  &
 $M_{\odot}$yr$^{-1}$ & km\,s$^{-1}$ &    &mag        &     \\
}
\startdata
NGC\,2363-V1 & Nov 1997 & 11    &590       & 6.35         & $-$3.1 & 325 &
4.0?& $-$10.4 & This work\\
& Jul 1999 & 13    &675       & 6.4          & $-$3.1 & 290 &
4.0?& $-$10.3 & This work\\
HD\,5980   & Sep 1994 & 23    &280       & 7.05          & $-$3.1 & 500 &0.7& $-$10.5    &This work\\
& Dec 1994 & 35    &130:     & 6.5          & $-$3.0 & 600 &2.5& $-$8.6    &(1)\\
$\eta$ Car & Mar 1998 & 27    &881      & 6.7          & $-$2.5 & 500 & 5?& $-$?    &(2)\\
\enddata
\tablecomments{We estimate that a volume filling
factor of $f$=0.3 is appropriate for V1, with $f$=0.1 for $\eta$ Car
and HD\,5980.
References: (1) Koenigsberger et al. (1998); (2) Hillier et al. (2000).}

\end{deluxetable}

\newpage

\begin{figure}
\figurenum{1}
\plotone{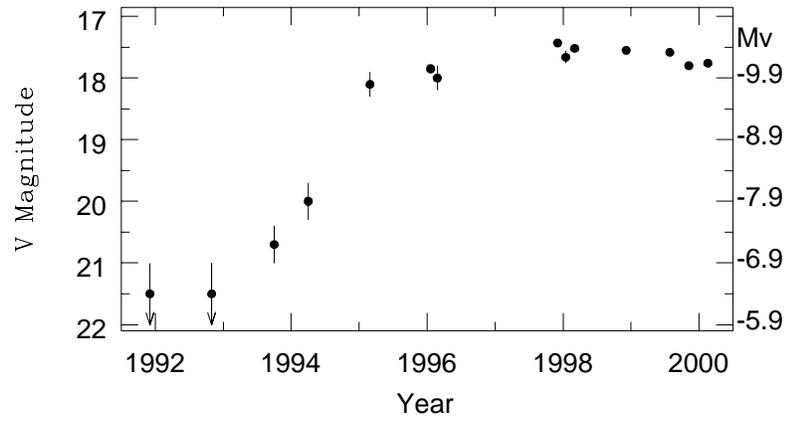}
\caption{V-band light curve of NGC\,2363-V1 over the past 7 years.
The absolute magnitude scale on the right assumes a distance 
of 3.44 Mpc and $E(B-V) = 0.06$ (see text).
}
\end{figure}

\clearpage

\begin{figure}
\figurenum{2}
\epsscale{0.8}
\plotone{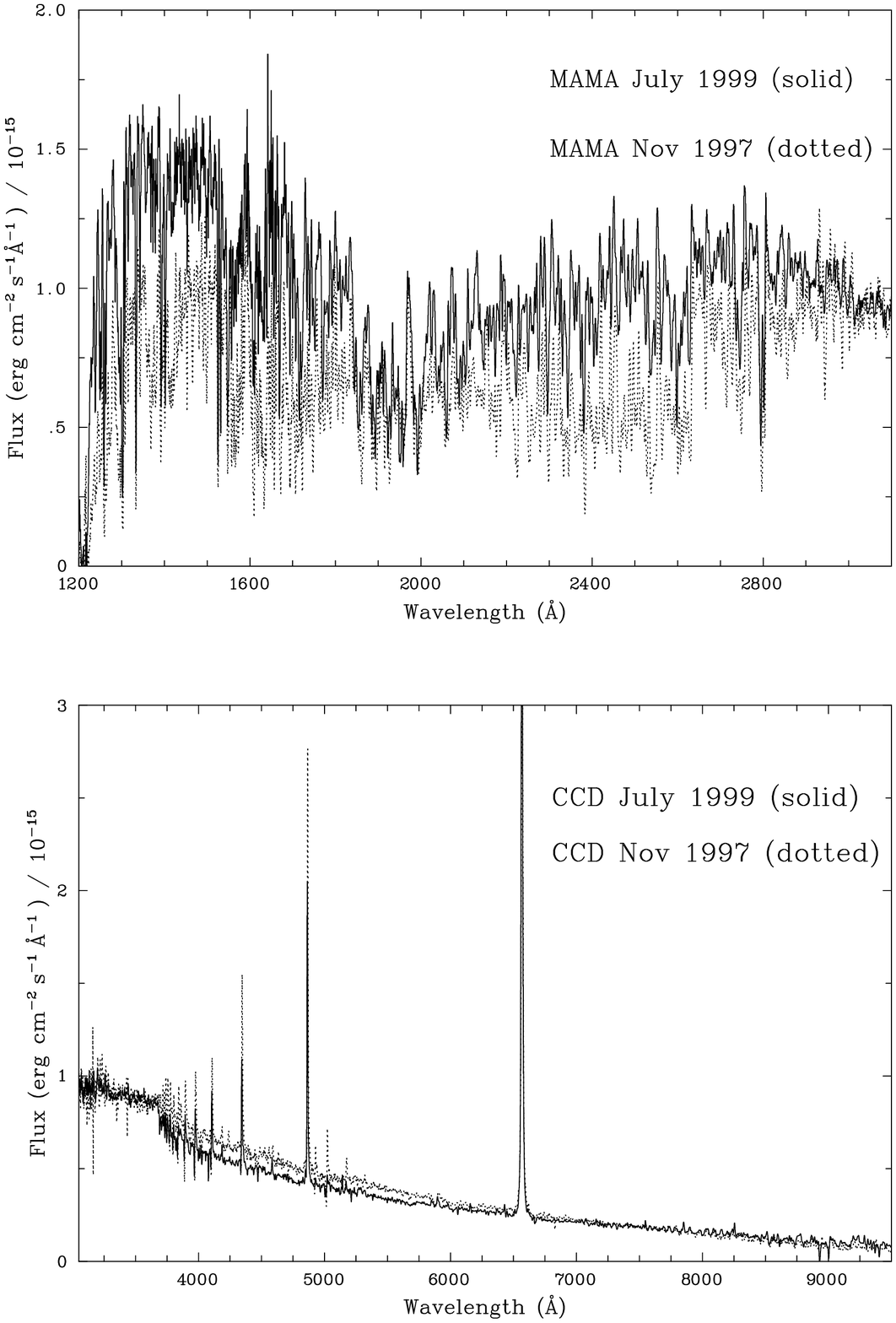}
\caption{Flux calibrated HST/STIS spectrogram of V1 from 1999 
July (solid) and 1997 November (dotted). The ripples longward of 7500 \AA\
are due to fringing on the CCD.
}
\end{figure}

\begin{figure}
\figurenum{3}
\plotone{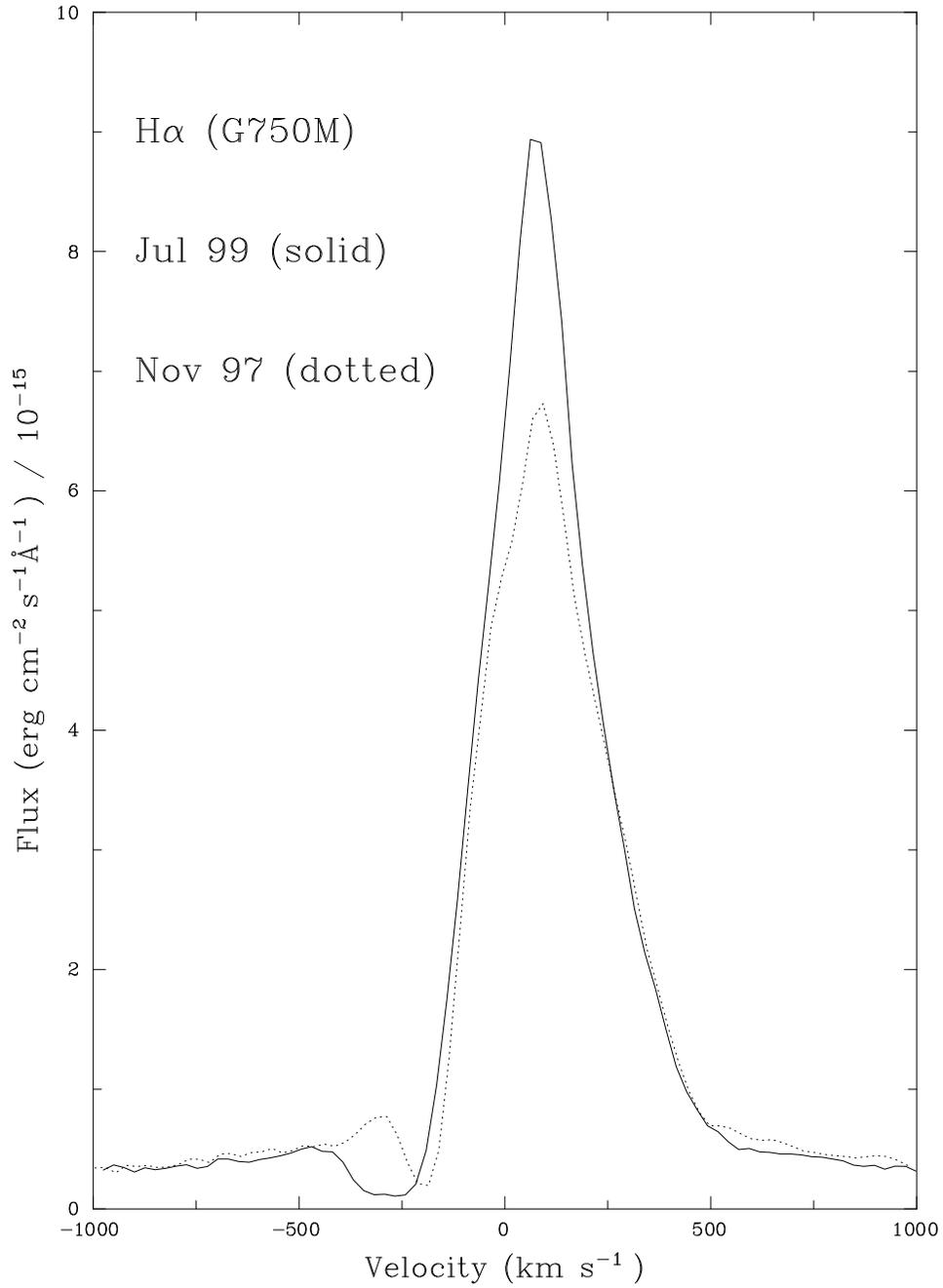}
\figcaption{High resolution HST/STIS spectrogram of V1 (grating
G750M), revealing a P Cygni H$\alpha$ profile.}
\end{figure}

\begin{figure}
\figurenum{4}
\plotone{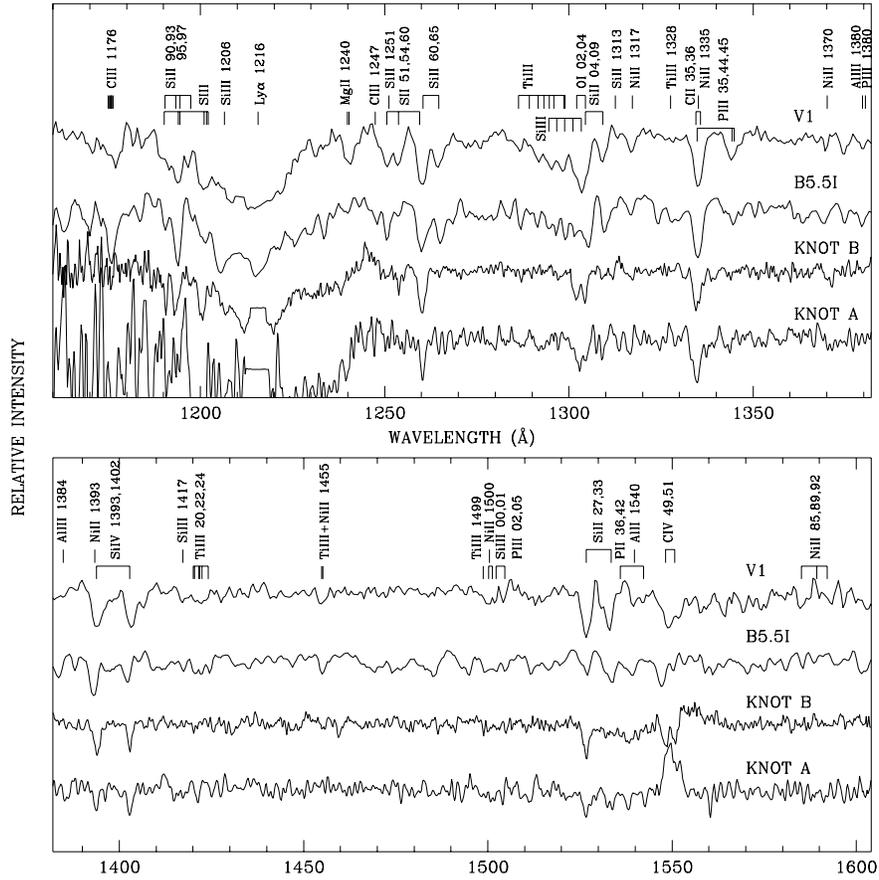}
\figcaption{Rectified spectra of NGC 2363-V1 in 1999, NGC 2363-A, NGC 2363-B and a
representative Galactic B5.5I spectrum (see text).}
\end{figure}

\begin{figure}
\figurenum{5}
\plotone{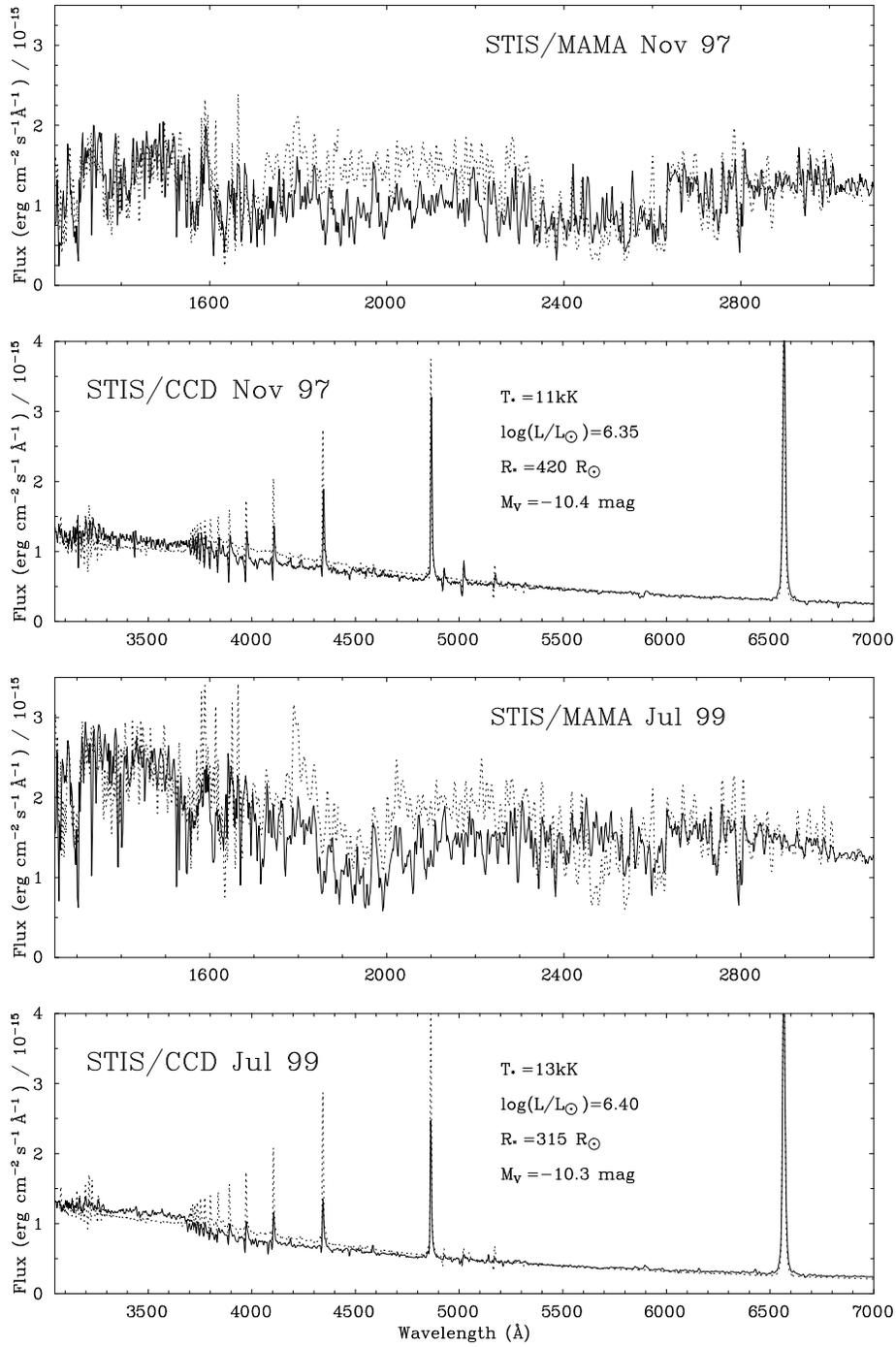}
\figcaption{Comparison between de-reddened 
HST/STIS datasets (solid) from 1997 November (upper panels) 
or 1999 July (lower panels) with non-LTE model predictions (dotted).}
\end{figure}

\begin{figure}
\figurenum{6}
\plotone{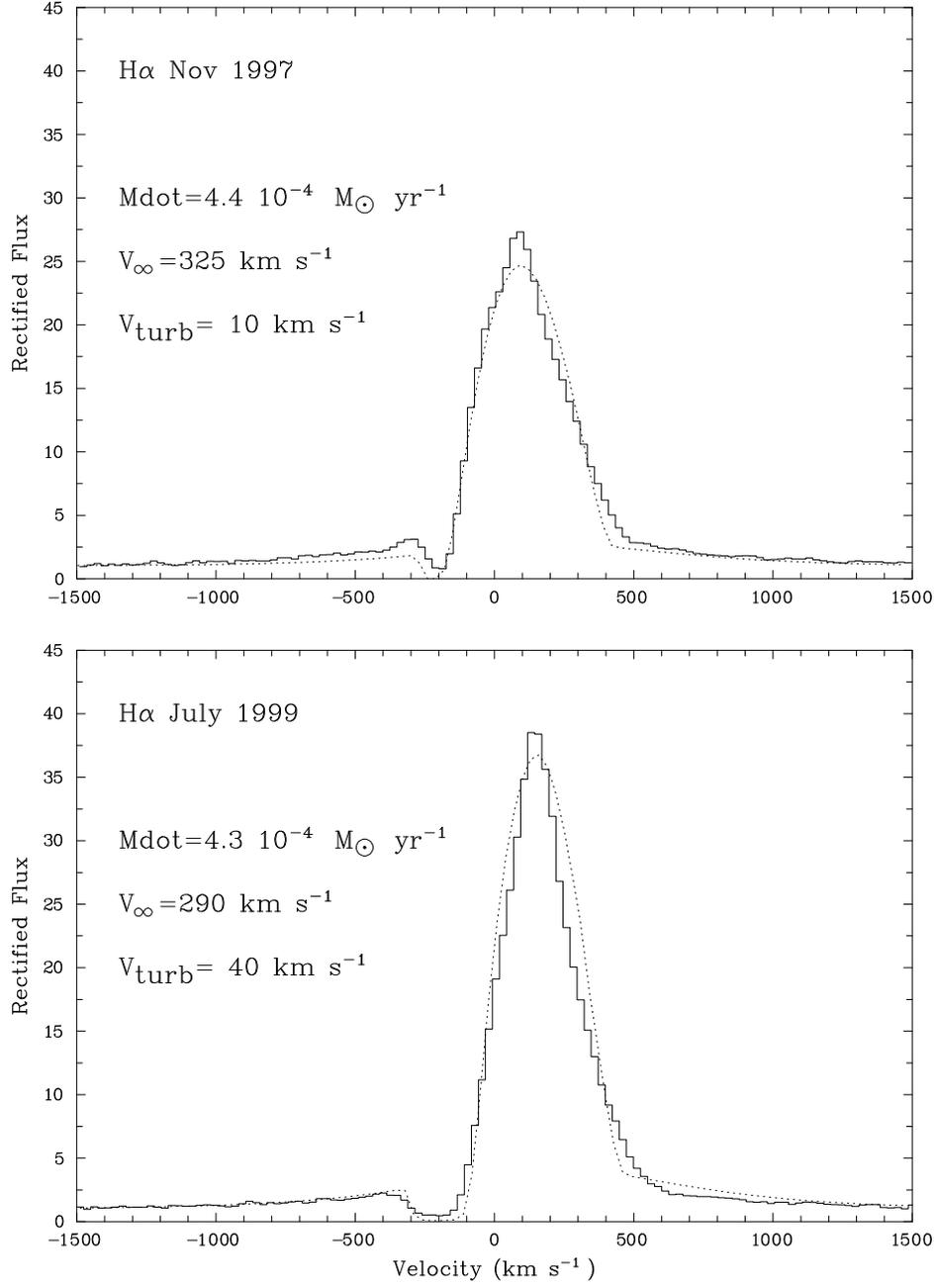}
\figcaption{Comparison between rectified H$\alpha$ profiles (solid)
and synthetic spectra (dotted).}
\end{figure}

\newpage

\begin{figure}
\figurenum{7}
\plotone{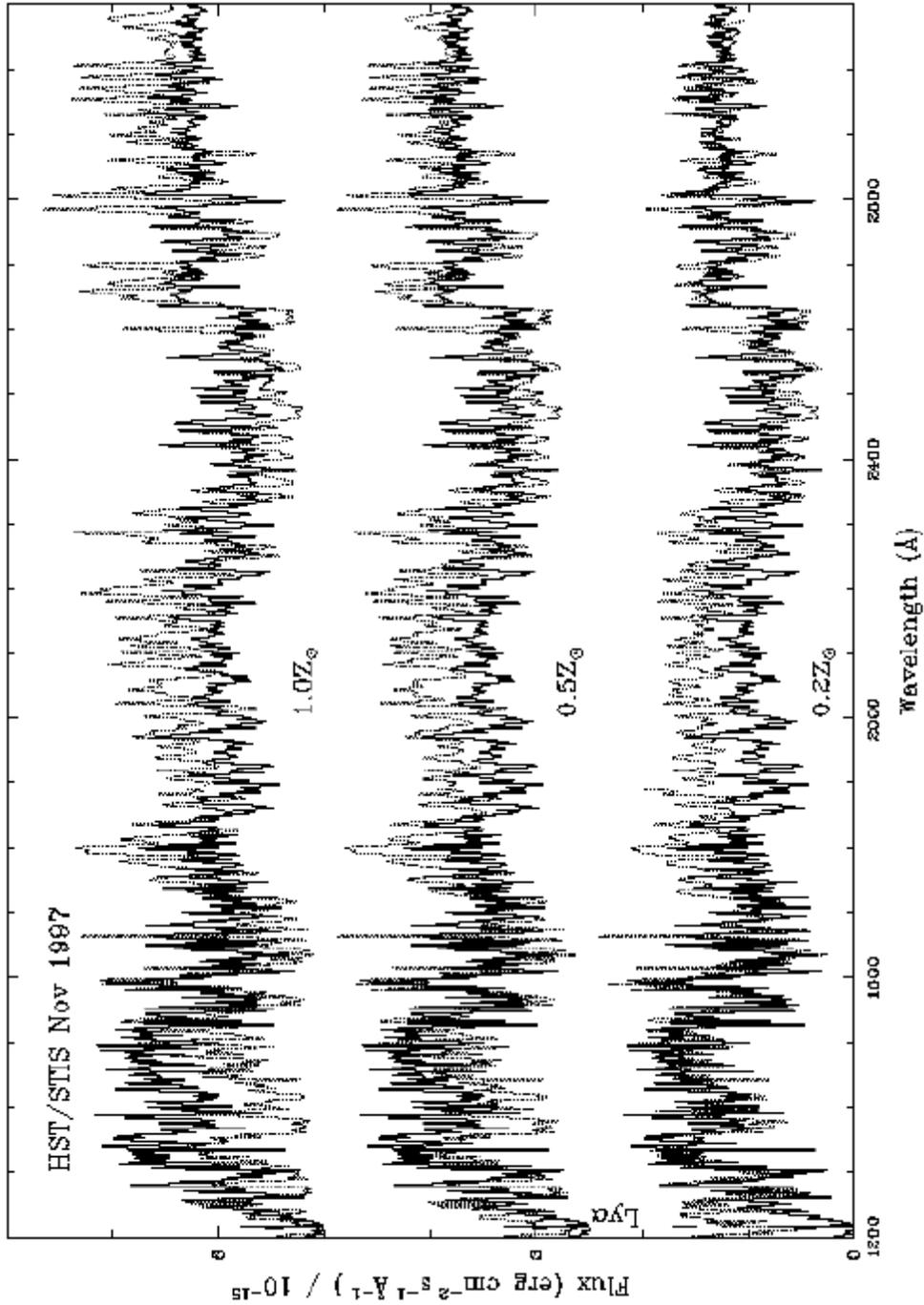}
\figcaption{Comparison between de-reddened STIS/MAMA datasets of V1
with line blanketed UV energy distributions, for 0.20$Z_{\odot}$ 
(lower), 0.5$Z_{\odot}$ (middle), 1.0$Z_{\odot}$ (upper), for clarity
successively offset by 2.5$\times$10$^{-15}$ erg\,cm$^{-2}$\,s$^{-1}$\,\AA.}
\end{figure}

\newpage
\begin{figure}
\figurenum{8}
\plotone{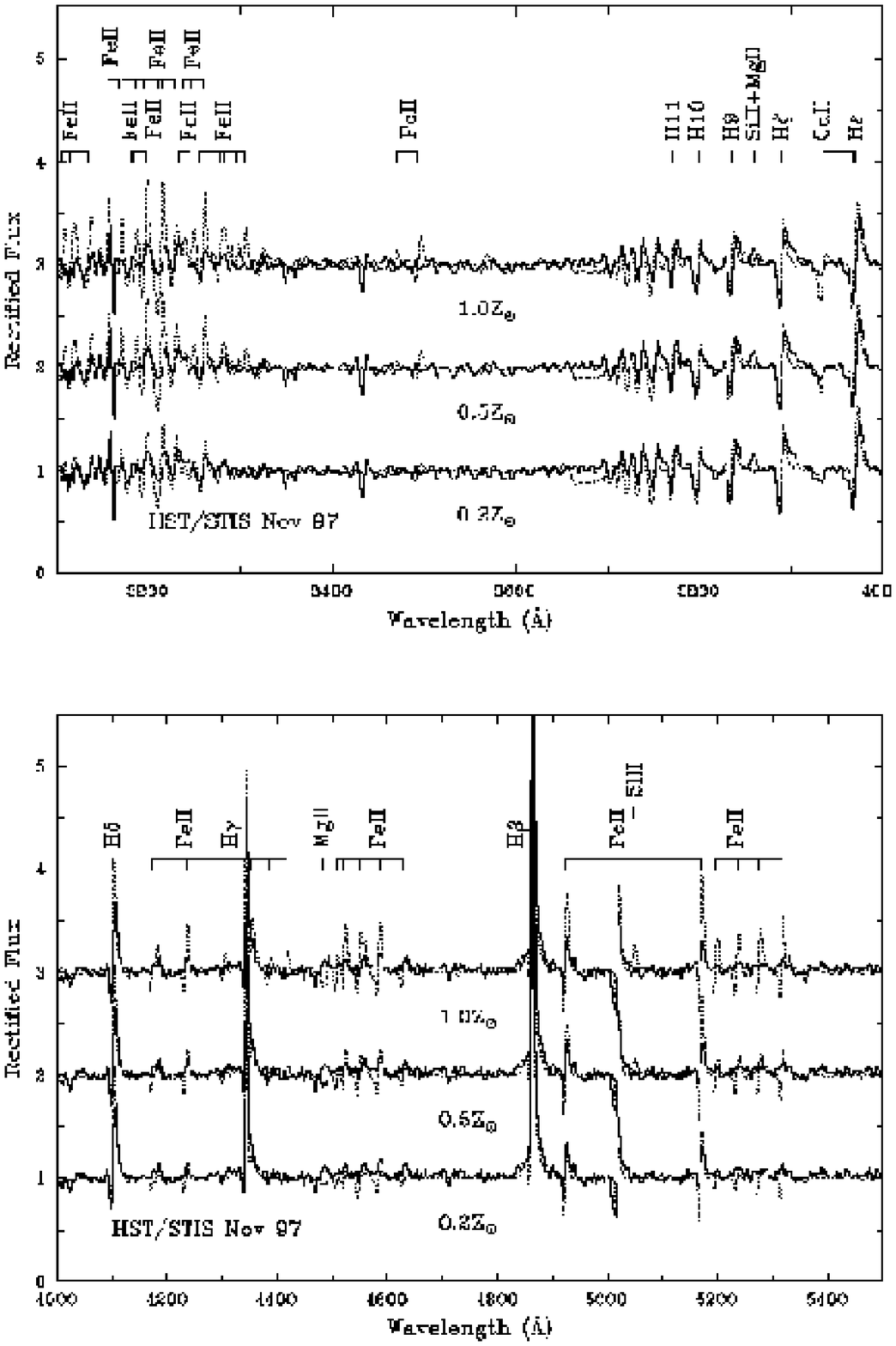}
\figcaption{Comparison between rectified, optical STIS datasets of 
V1 with line blanketed energy distributions, for 
0.20$Z_{\odot}$ (lower), 0.5$Z_{\odot}$ (middle), 1.0$Z_{\odot}$ (upper),
for clarity successively offset by a continuum unit.}
\end{figure}

\newpage

\begin{figure}
\figurenum{9}
\plotone{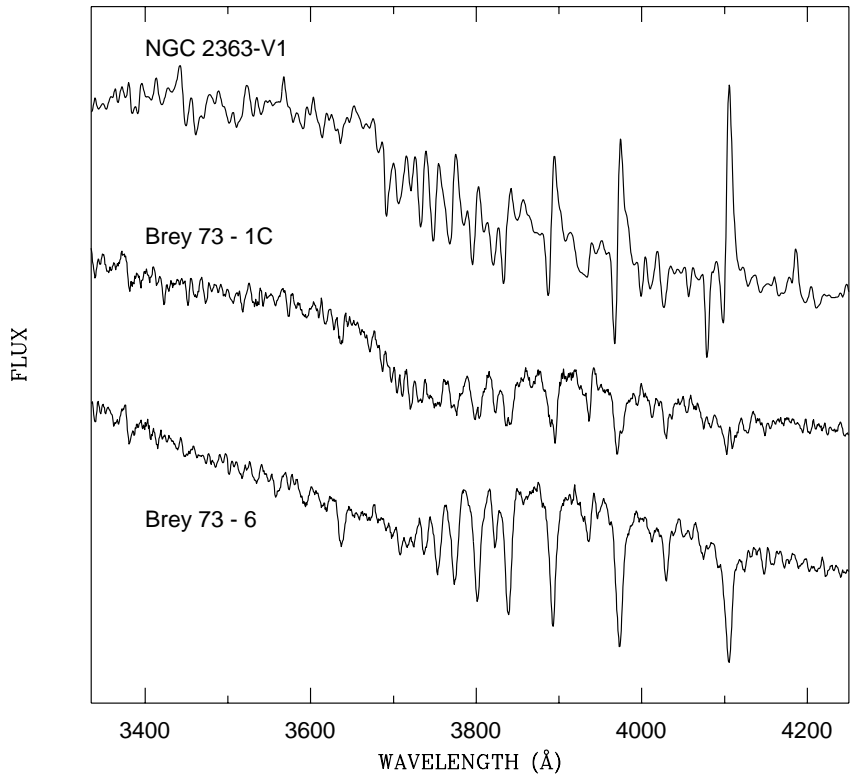}
\figcaption{Comparison between V1 and two LMC B stars in the 
Breysacher~73 cluster, \# 1C (O9.5-B1pe) and \# 6 (B0.2V), in
the vicinity of the Balmer jump.}
\end{figure}

\begin{figure}
\figurenum{10}
\plotone{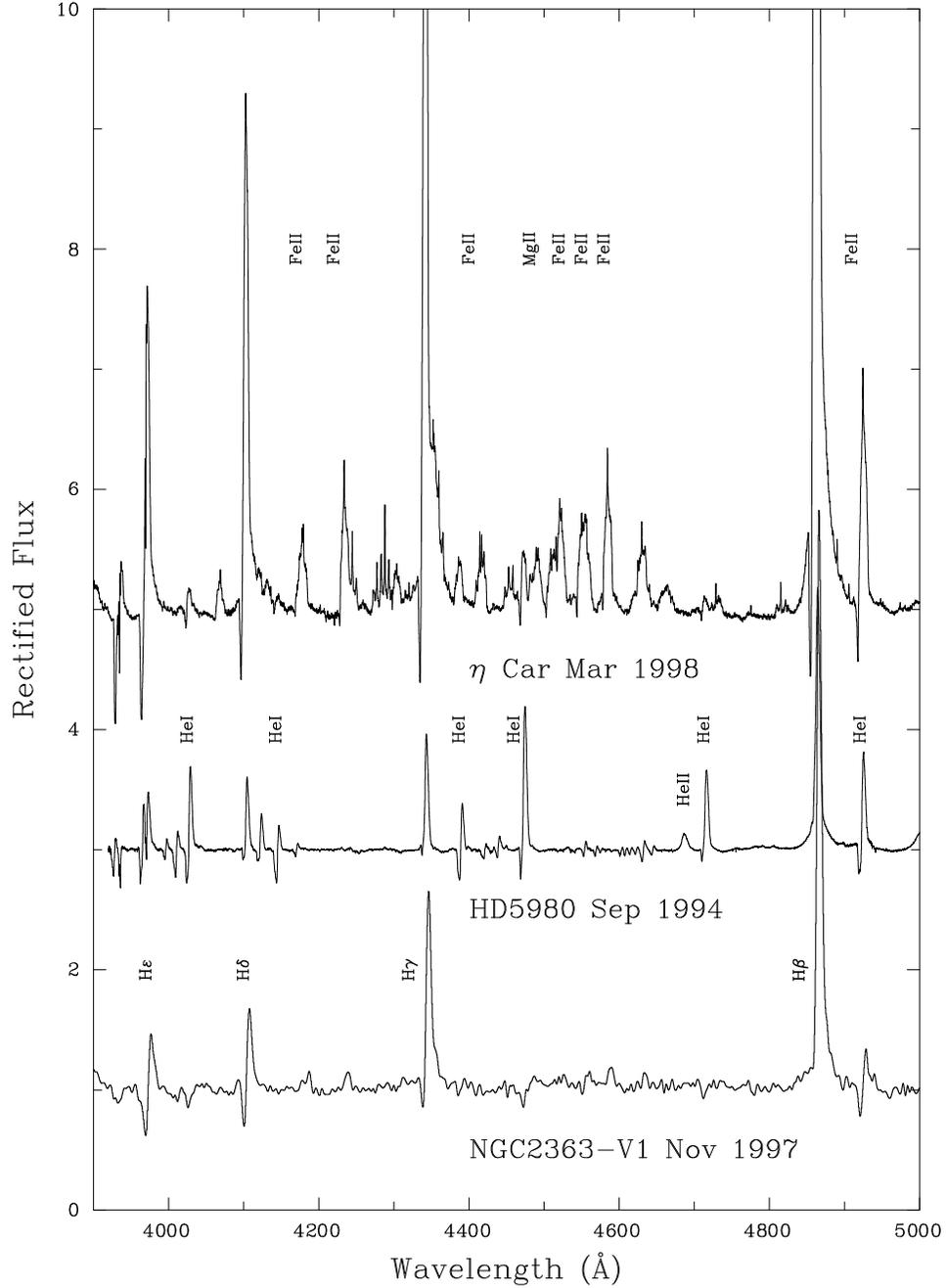}
\figcaption{Comparison between rectified, optical datasets of 
V1 with HD\,5980 during outburst in 1994 September, plus $\eta$ Car
in 1998 March (HST/STIS data from Davidson \et, in preparation).}
\end{figure}

\end{document}